\documentclass[aps, prd, superscriptaddress, preprintnumbers, floatfix, longbibliography, 10pt]{revtex4-2}

\usepackage{graphicx,amsfonts,amsmath,amssymb,amstext}
\usepackage{float,wrapfig}
\usepackage{subfigure, psfrag}
\usepackage{dsfont}
\usepackage{xcolor}
\usepackage[colorlinks=true,urlcolor=blue,citecolor=blue,linkcolor=blue]{hyperref}
\usepackage{bm}
\usepackage{mathtools}
\usepackage[normalem]{ulem}
\usepackage{adjustbox}

\graphicspath{{Figures/}} 

\definecolor{purple}{rgb}{0.8,0,0.6}
\definecolor{darkgreen}{rgb}{0.00,0.6,0.00}

\newcommand{\revisionA}[1]{\textcolor{blue}{#1}}

\extrafloats{100}

\begin{document}

\title{Electron binding energy of a donor in bilayer graphene with a gate-tunable gap}

\author{E. V. Gorbar}
\affiliation{Department of Physics, Taras Shevchenko National University of Kyiv, Kyiv, 03022, Ukraine}
\affiliation{Bogolyubov Institute for Theoretical Physics, Kyiv, 03143, Ukraine}

\author{V. P. Gusynin}
\affiliation{Bogolyubov Institute for Theoretical Physics, Kyiv, 03143, Ukraine}

\author{D. O. Oriekhov}
\affiliation{Instituut-Lorentz, Universiteit Leiden, P.O. Box 9506, 2300 RA Leiden, The Netherlands}
\affiliation{Kavli Institute of Nanoscience, Delft University of Technology, 2628 CJ Delft, the Netherlands}

\author{B. I. Shklovskii}
\affiliation{School of Physics and Astronomy, University of Minnesota, Minneapolis, Minnesota 55455, USA}

\begin{abstract}

In gapped bilayer graphene, similarly to conventional semiconductors, Coulomb impurities (such as nitrogen donors) may determine the activation energy of its conductivity and provide low temperature hopping conductivity. However, in spite of importance of Coulomb impurities, nothing is known about their electron binding energy $E_b$ in the presence of gates. To close this gap, we study numerically the electron binding energy $E_b$ of a singly charged donor in BN-enveloped bilayer graphene with the top and bottom gates at distance $d$ and gate-tunable gap $2\Delta$. We show that for $10 < d < 200$\,\text{nm} and $1 < \Delta < 70$\,\text{meV} the ratio $E_b/\Delta$ changes from 0.4 to 1.4. The ratio $E_b/\Delta$ stays so close to unity because of the dominating role of the bilayer polarization screening which reduces the Coulomb potential well depth to values $\sim \Delta$. Still the ratio $E_b/\Delta$ somewhat decreases with growing $\Delta$, faster at small $\Delta$ and slower at large $\Delta$. On the other hand, $E_b/\Delta$ weakly grows with $d$, again faster at small $\Delta$ and slower at large $\Delta$. We also studied the effect of trigonal warping and found only a small reduction of $E_b/\Delta$.

\end{abstract}

\maketitle

\section{Introduction}
\vspace{5mm}

Bernal-stacked bilayer graphene (BLG) is composed of two layers of graphene
where atoms of sublattice A in one layer lie directly below or above atoms of sublattice B in the other layer so that interlayer dimer bonds are formed. Its low energy electron spectrum is gapless and is given by parabolic valence and conduction bands touching at two $K$ and $K^{\prime}$ valley points. Their charge carriers are characterized by non-trivial chiral properties with Berry phase $2\pi$ \cite{McCann, Novoselov} and are described by two-component spinors. An unique feature of BLG is that an electric field $D$ applied perpendicular to its layers due to potential difference between gates opens gap $2\Delta$ between the valence and conduction bands~\cite{McCann, Novoselov, Neto, Castro}. The possibility of inducing and controlling the energy gap via gating makes BLG an active research area including practical applications in electronic devices. We assume below that as in the best devices BLG is enveloped by two layers of hexagonal boron nitride (hBN) with thickness $d$ each and has top and bottom gates made of graphite or gold with electric potential $+U$ and $-U$, respectively, (see Fig.1a). This potential difference opens gap $2\Delta=-2eUc_0/d$ in bilayer graphene, where $c_0 \approx 0.34\,\text{nm}$ is interlayer separation and $-e<0$ is the electron charge.

\begin{figure}
    \centering
\includegraphics[scale=0.5]{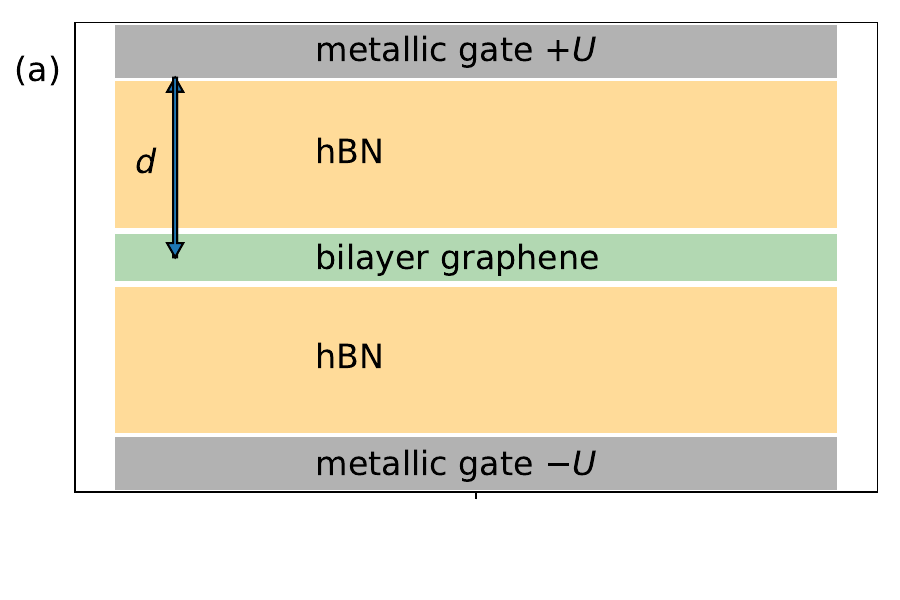}
 \includegraphics[scale=0.5]{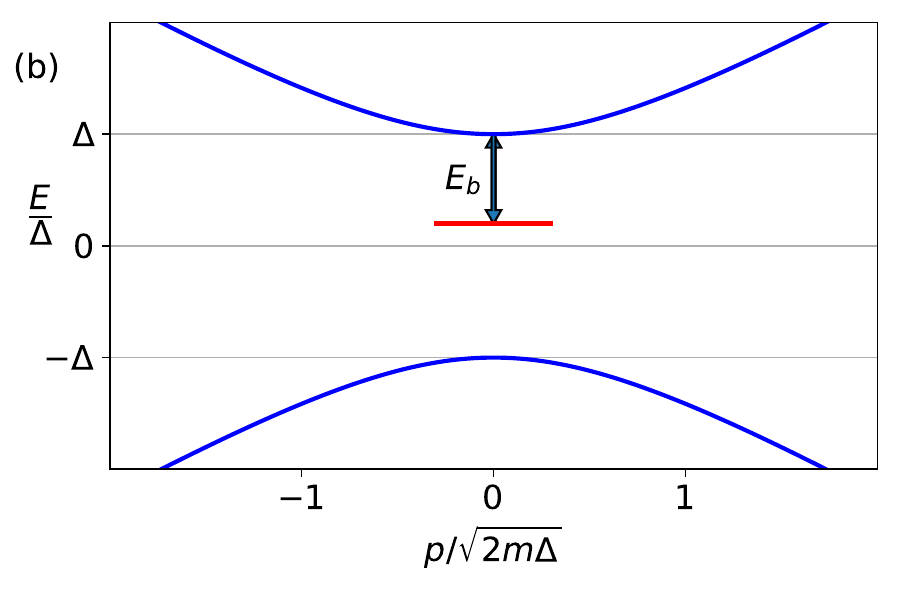}
\caption{(a) Schematic geometry of dual gated device studied. hBN substrates are characterized by dielectric constant $\kappa$ and $d$ denotes the distance to gates. (b) The energy dispersion of the two-band model of bilayer graphene without trigonal warping and the definition of binding energy $E_b$ of an in-gap bound state.}
    \label{fig:intro-parameters}
\end{figure}

In this paper we study the ground state of the electron bound to singly charged positive donor. It is known that best samples of graphite and, consequently, exfoliated from it BLG samples have charged impurities in concentration $\sim 10^{9}$ cm$^{-2}$ per layer, presumably singly charged nitrogen donors~\cite{Young2017, Joucken2021,Young2022}. These donors provide activated electron conductivity with activation energy related to their binding (ionization) energy $E_b$. At low temperature donors compensated by smaller concentration of acceptors provide weakly activated hopping conductivity. Donors are also responsible for additional lines of the light absorption. Thus, it is important to find the donor ground state energy $E(\Delta, d)$ calculated from the mid-gap energy $E = 0$ and electron binding (ionization) energy $E_b = \Delta - E$ (see Fig.\ref{fig:intro-parameters}(b)). History of semiconductors implies that knowledge of binding energy of donors and acceptors is important both for efforts to make BLG cleaner or for intentional doping of BLG, which eventually allow to create better devices.

The parabolic energy spectrum in BLG plotted in Fig.1b is actually realized only at low-energy energy \cite{McCann} for momenta up to $|\mathbf{p}| \approx \gamma_1/(2v_F)$, where $v_F$ is the Fermi velocity and $\gamma_1$ is the interlayer hopping amplitude.
The energy dispersion is linear for larger momenta as in single layer graphene. The four-band model formulated in \cite{McCann} describes the electron quasiparticles in BLG both at low and high energies. Its energy dispersion in gapped bilayer graphene has a Mexican hat structure where the smallest by modulus value of energy is realized at nonzero momentum $\mathbf{p}_0$ rather than at $\mathbf{p}=0$. Still such a Mexican hat structure is negligible at small gap $\Delta$ because the difference of energies at zero momentum and $\mathbf{p}_0$ is proportional to $2\Delta^3/\gamma^2_1$. Therefore, it is quite small at $\Delta \ll \gamma_1$. In addition, momentum $|\mathbf{p}_0| \approx \sqrt{2}\Delta/v_F$ tends to zero as $\Delta \to 0$. Thus, the Mexican hat structure is smoothed and shrinks to zero momentum for small gap. Since we consider in our paper small gaps, we employ the two-band model whose energy dispersion increases monotonously with momentum and does not have the Mexican hat shape. This model is applicable for gap $\Delta < \gamma_1/4 \approx 100\,\text{meV}$. Our calculations are limited by $\Delta= 70\,\text{meV}$.


Our calculation takes into account two kinds of screening of the Coulomb potential of a donor. First, we take care of polarization of gapped BLG via virtual electron transitions across the gap~\cite{Silvestrov,bilayer-instability}, as well as polarization due to $\sigma$-bonds electrons \cite{Ulybyshev,Santos}. This screening is very strong at small $\Delta$ and as a result reduces the Coulomb potential of a donor to a shallow long range potential well with the depth of order $\Delta$~\cite{Silvestrov}. Note that this gapped BLG polarization screening is different from screening of BLG without gap~\cite{Gamayun2011PRB,Nandkishore2010PRL_exciton,Nandkishore2010PRB, Nandkishore2010PRB2}. Second, there is screening due to top and bottom gates. It is known that screening of a single charge located in the middle between two metallic gates (in the absence of a BLG polarization) leads to the exponential decay of potential with the decay length $2d/\pi$~\cite{Pumplin,Glasser}. In our case, BLG releases electric lines to the gates only at large distances from the donor. Therefore, as we see below, the effect of screening by gates in the double screened potential is quite modest. 

In our calculations, we use techniques developed in the previous study of electron bound states of the Coulomb center with an arbitrary charge $Z$ in BLG~\cite{bilayer-instability}. That work was mostly concerned with finding a critical $Z_c$ at which energy of the ground state donor level $E$ reaches $-\Delta$ and dives into the valence band. Besides that, Ref.\cite{bilayer-instability} did not study screening by gates. The main objective of the present work is to find binding energy of donor $E_b$ with single positive charge as a function of $\Delta$ and $d$ taking into into account screening of the potential of charged impurity both by gates and BLG electrons.

\begin{figure}[ht]
\includegraphics[scale=0.5]{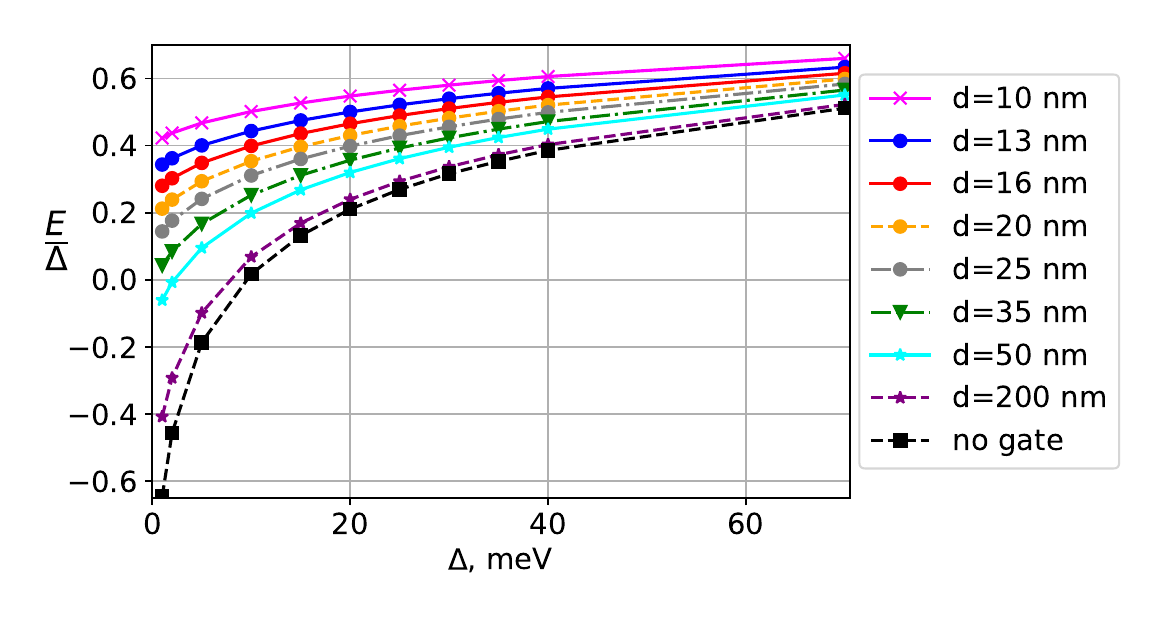}
\caption{The ground state energy of donor $E$ as a function of $\Delta$ in units of $\Delta$ for few values of distance to gates $d$ from $10$ to $200\,\text{nm}$ as well as in the absence of gates.}
\label{exact-levels-gated-potential}
\end{figure}

In Fig. \ref{exact-levels-gated-potential}, we present our results for $E(\Delta, d)$ with $10 < d < 200$\,\text{nm} and $1 < \Delta < 70$\,\text{meV} taking into account double screening due to gates and bilayer graphene including $\sigma$-bonds polarization. We see that the ratio $E_b/\Delta$ changes from 0.4 to 1.4. It stays close to unity because of the dominating role of the BLG polarization screening which reduces the Coulomb potential well depth $\sim \Delta$~\cite{Silvestrov}. Still the ratio $E_b/\Delta$ somewhat decreases with growing $\Delta$, faster at small $\Delta$ and slower at large $\Delta$. On the other hand, $E_b/\Delta$ grows with $d$ faster at small $\Delta$ and slower at large $\Delta$. This happens because at small $\Delta$ the localization length of the electron bound state becomes larger and screening by gates plays larger role.  
We also studied the effect of the trigonal warping at $\Delta \leq 20\,\text{meV}$ and found only a small reduction of $E_b/\Delta$.

The paper is organized as follows. The low-energy effective Hamiltonian, the equations of motion, and numerical results for the donor binding energy in the two-band model are presented in Sec.\ref{sec:two-band-model}. The role of trigonal warping effects on the electron bound states is described in Sec.\ref{sec:trigonal-warping}.  The electric potential of Coulomb impurity screened both by internal electron polarization and by gates in gated BLG is found in Sec.\ref{sec:electric-potential}.
Main results and conclusions are summarized in Sec.\ref{sec:conclusions}. An approximate analytic method for estimation of the electron bound state energy is presented in Appendix
\ref{sec:appendix-A}.

\section{Hamiltonian, equations of motion, and numerical results}
\label{sec:two-band-model}

\vspace{5mm}

The free two-band Hamiltonian in gapped bilayer graphene has the form \cite{McCann}
\begin{equation}
  H_0=\frac{v^2_F}{\gamma_1}\left(\begin{array}{cc}
  0&(p_{-})^2\\
  (p_{+})^2&0
  \end{array}\right)+\Delta\left(\begin{array}{cc}
  1&0\\
  0&-1
  \end{array}\right),
  \label{Hamiltonian}
  \end{equation}
where $p_\pm=p_x\pm i p_y$ and ${\bf p}=-i \hbar\bm{\nabla}$ is the two-dimensional momentum operator,
$\gamma_1 \approx 0.39~\mbox{eV}$ \cite{McCann2007gamma1} is the strongest interlayer coupling between the pairs of orbitals that
lie directly below and above each other, and $v_F
\simeq c/300$ is the Fermi velocity. The two component spinor
field $\Psi_{Vs}$ carries the valley $(V= K, K^{\prime})$ and spin
$(s = +, -)$ indices. We will use the standard convention:
$\Psi_{Ks}^T=(\psi_A{_1}, \psi_B{_2})_{Ks}$
whereas $\Psi_{K^{\prime}s}^T = (\psi_B{_2},
\psi_A{_1})_{K^{\prime}s}$. Here $A_1$ and $B_2$ correspond to
those sublattices in layers 1 and 2, respectively, which, according to the
Bernal $(A_2-B_1)$ stacking,
are relevant for the low energy dynamics.

The energy spectrum of Hamiltonian (\ref{Hamiltonian}) is given by
$E({\bf p})=\pm\sqrt{({\bf p}^2/2m)^2+\Delta^2}$, so that we have gap $2\Delta$ between the lower
and upper continua (valence and conduction bands), where $m=\gamma_1/2v_F^2\approx0.054 m_e$ is
the quasiparticle mass and $m_e$ is the mass of the electron.

To determine the electron bound states in the field of Coulomb impurity in gated BLG we should find the electric potential $V(r)$ induced by this impurity.
We solve the Poisson equation taking into account the polarization effects in BLG and boundary conditions at gates and obtain the corresponding electric potential
in Sec.\ref{sec:electric-potential}.

The exact calculation of energy levels proceeds employing the integral equations in momentum space following the approach used in Ref.\cite{bilayer-instability}. In numerical analysis, it is convenient to
represent the integral equations for the wave function of electron bound states in the field of Coulomb impurity in dimensionless variables. Using the wave function in the form
\begin{align}
	\Psi=\left(\begin{array}{l}
		 e^{i(j-1) \theta}a_j(k) \\
		 e^{i(j+1) \theta}b_j(k)
	\end{array}\right),
 \label{wavefunction}
\end{align}
we find that the corresponding integral
equations for two components of the electron spinor function with the total angular momentum $j=0,\pm 1,\pm 2,...$ in momentum space have the form
\begin{align}\label{eq:system-potential-1}
&k^2b_j(k)+a_j(k)-\xi\int^{Q}_0dp\,pa_j(p)\,V_{j-1}(k,p)=\epsilon a_j(k),\\
\label{eq:system-potential-2}
&k^2a_j(k)-b_j(k)-\xi\int^{Q}_0dp\,pb_j(p)\,V_{j+1}(k,p)=\epsilon b_j(k),
\end{align}
where $\xi=(\alpha_g/\kappa)(\gamma_1/\Delta)^{1/2}$,
$\alpha_g=e^2/(\hbar v_F)\approx 2.19$ is the fine structure constant in graphene, energies are expressed in units of gap $\Delta$, and momenta are rescaled by the inverse of the wave length
$\lambda_1=\hbar v_F / \sqrt{\gamma_1 \Delta}$.
The cut-off parameter $Q=\sqrt{\gamma_1/\Delta}$ corresponds to the range of applicability of the effective 
model \eqref{Hamiltonian}. To get the above set of 1D integral equations, we integrated the potential over angle
\begin{align}\label{eq:V_j-definition}
	V_j(k, p)=\frac{1}{2\pi} \int_0^{2 \pi} d \varphi\,\tilde{\phi}\left(\sqrt{k^2+p^2-2 k p \cos \varphi}\right) \cos (j \varphi),\quad \phi(|\mathbf{k}|)=(2\pi e \lambda_1/\kappa)\tilde\phi(|\mathbf{k}|\lambda_1),
\end{align}
where the dimensionless potential $\tilde{\phi}(p)$ is expressed through the screened potential $\phi(|\mathbf{k}|)$ in Eq.\eqref{potential-with-gates} with charge $q=e$. The numerical solution is found in the same way as in Ref.\cite{bilayer-instability} via discretization of momentum in a grid of $N$ points
$k=i Q / N,\,i=1,2,\dots$ and replacing integrals by summation $\int_{0}^{Q}dk f(k)\to \sum_{i=1}^{N} w_i f(iQ/ N)$
with weight coefficients $w_i$ of discrete quadrature formulas. The Newton-Cotes formulas of the fifth order are used at each sub-intervals of 6 neighboring points to determine the weight coefficients $w_{i}$. 
Such an approach reduces the discrete integration error to the order of $h^9 f^{(8)}(q)$, where $h$ is the integration step and $f^{(8)}(q)$ is the 8-th derivative of the integrand. Since the wave functions are expected to be regular functions in the considered region, the integration error quickly reduces with decreasing step size. It is important to note that the kernel functions $V_{j}(k,p)$ in the presence of polarization and gating are not singular at $k=p$. Therefore, the technique used in Ref.\cite{bilayer-instability} to extract and regularize the singularity coming from the Coulomb potential is not needed here. For each pair of momentum points $k_i, p_i$ we evaluate numerically the integral over angle in Eq.\eqref{eq:V_j-definition} with potential \eqref{potential-with-gates}. After such discretization of momentum space integration,  the system of equations \eqref{eq:system-potential-1} and 
\eqref{eq:system-potential-2} becomes a matrix equation for $2N$ unknowns whose eigenvalues define the spectrum. A technical remark regarding this solution should be 
made. The discrete grid starts not exactly at zero momentum to avoid the appearance of a spurious solution with wave function fully localized at zero momentum. The 
convergence is ensured by taking a sufficiently large number of points $N$, which equals $N=1000$ for the smallest gap value considered. 

\begin{figure}[ht]
\includegraphics[scale=0.5]{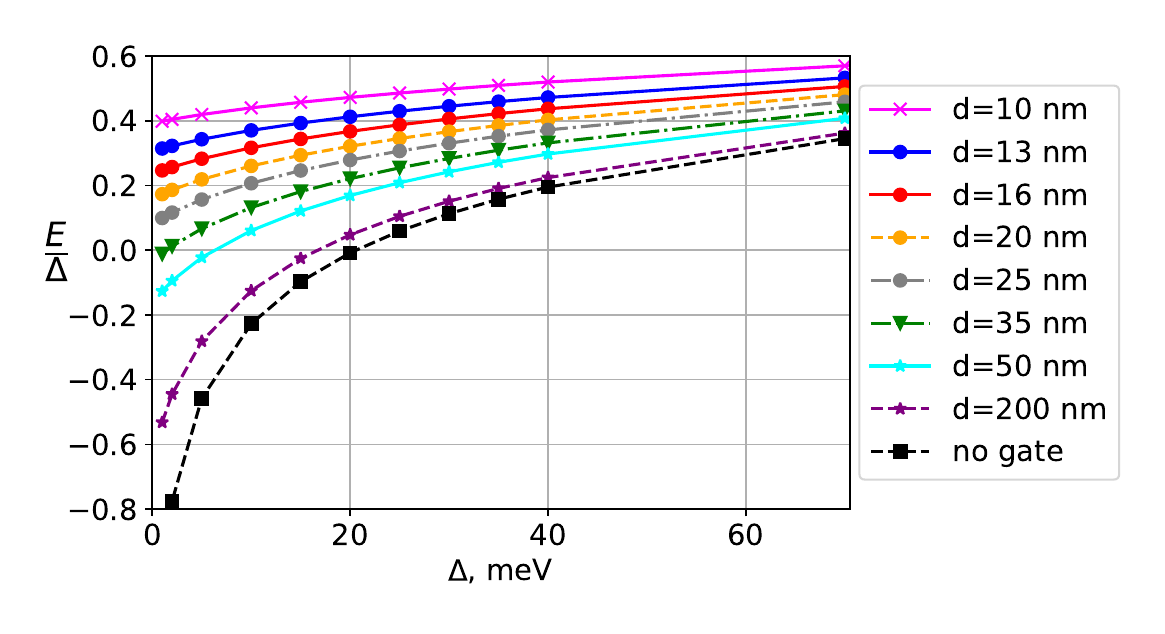}
\caption{The ground state energy of donor $E$ as a function of $\Delta$ in units of $\Delta$ for few values of distance to gates $d$ from $10$ to $200\,\text{nm}$ in the absence of screening due to $\sigma$-bonds electrons.}
\label{exact-levels-gated-potential-absence}
\end{figure}

The results are presented in Fig.\ref{exact-levels-gated-potential} where the energy of the lowest energy state in units $\Delta$ is plotted for $q=e$ and $\kappa=4$ as a function of
$\Delta$ for few values of $d$ from $10$ to $200\,\text{nm}$ as well in the absence of gates. Analogous results are presented in Fig.\ref{exact-levels-gated-potential-absence} for the case when $\sigma$-bonds polarization is not taken into account. Comparing Figs. \ref{exact-levels-gated-potential} and \ref{exact-levels-gated-potential-absence} we conclude that screening due to $\sigma$-bonds electrons weakly affects the bound electron state energy for small gaps and distances (see, Table \ref{table-of-energies}).

Obviously, the influence of gates is the most significant at 
small gaps and strongly changes the relative energy position of the bound state. For convenience, the numerical values of $E$ for gaps $1\,\text{meV}$ and $2\,\text{meV}$ 
and several values of distance $d$ are given in Table \ref{table-of-energies} including (the last two rows) and in the absence of the contribution to the polarization function due to $\sigma$-bonds electrons (the 2nd and 3rd rows). Clearly, the correction to the electron bound state energy due to the $\sigma$-bonds polarization increases as $\Delta$ and $d$ grow.

For the considered values of $\Delta$, the radius of the donor wave function in the BLG plane is given by the effective Bohr radius $a_B =\hbar v_F/(\Delta \alpha_g)$ and exceeds $10$ nm. Such a radius is many times larger than the lattice constant of BLG $a_0 \sim 0.25$ nm. This means that our donor energies are valid for all singly charged impurities. Indeed, such impurities differ only in the chemical shift related to the short range core of impurity potential depending on the chemical nature of the atom. The role of the chemical shift is known to be very small when $a_B \gg a_0$ \cite{PhysRev.98.915,KOHN1957257,Bassani1974,Shklovskii2013}.

\begin{table}[]
\begin{tabular}{|l|l|l|l|l|l|l|l|}
\hline 
$\Delta$ \slash $\,\,d, nm$ & 10   & 13    & 16    & 20    & 25    & 35     & 50     \\
\hline
1 meV & 0.40  & 0.31 & 0.25 & 0.17 & 0.10   & -0.01 & -0.13 \\
\hline 
2 meV & 0.81 & 0.65 & 0.51 & 0.37 & 0.23 & 0.02 & -0.19\\
\hline
1 meV, $\sigma$-bonds polarization & 0.42  & 0.34 & 0.28 & 0.21 & 0.14   & 0.04 & -0.06 \\
\hline 
2 meV, $\sigma$-bonds polarization & 0.88 & 0.72 & 0.60 & 0.48 & 0.36 & 0.16 & -0.015\\
\hline
\end{tabular}
\caption{Numerical values of the lowest bound state energy $E$ in meV for different gap values $\Delta$ and distance to gates $d$.}
\label{table-of-energies}
\end{table}



\begin{figure}
    \centering
    \includegraphics[scale=0.5]{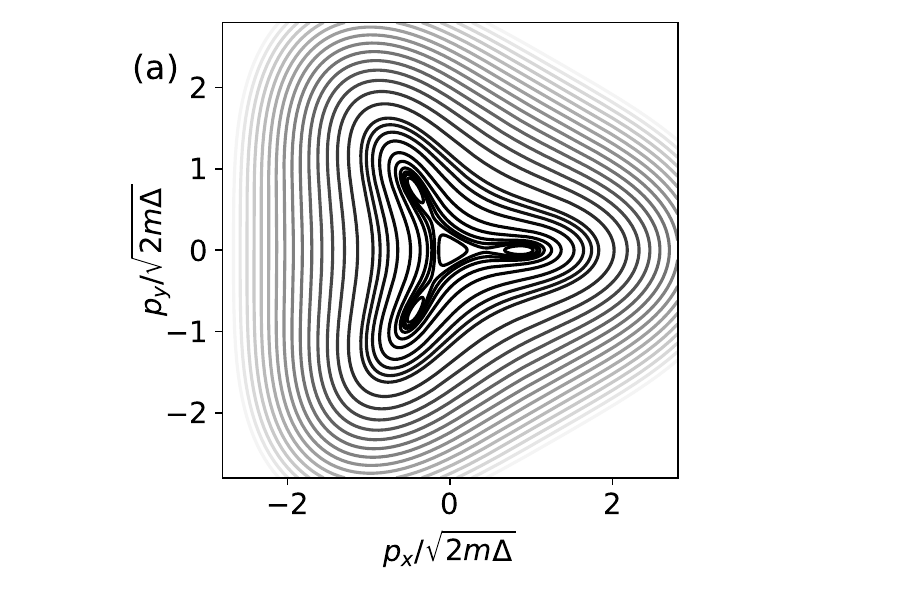}
    \includegraphics[scale=0.5]{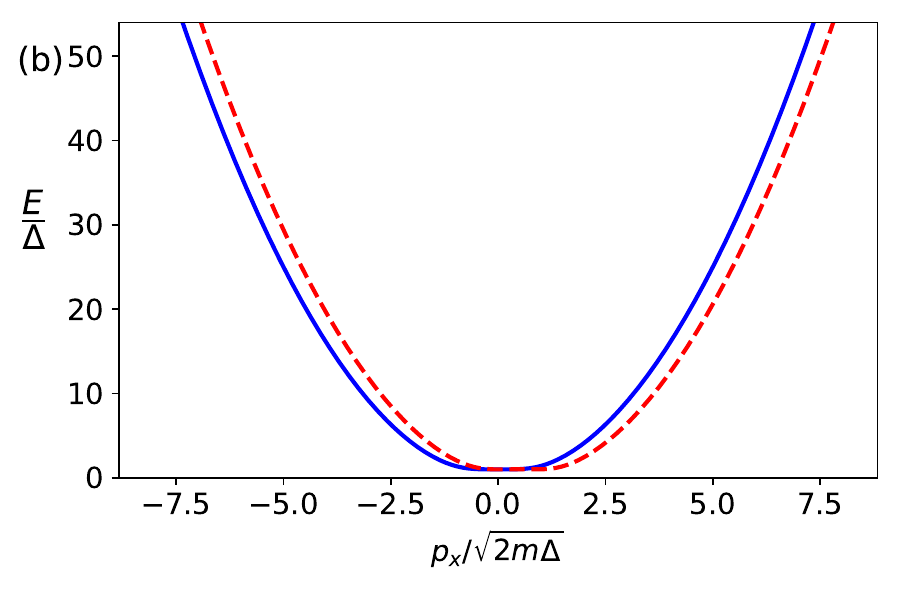}
    \caption{(a) A schematic view of the energy spectrum with trigonal warping corrections plotted for gap $\Delta=5$ meV. Contour lines corresponds to lines of constant energy. Dark contour lines enclose four domains where the minimal energy is realized at $E=\Delta$. (b) The energy spectrum at $p_y=0$ without (blue solid line) and with trigonal warping corrections (red dashed line).}
    \label{fig:trigonal-warping}
\end{figure}

\section{Trigonal warping effects}
\label{sec:trigonal-warping}

\vskip5mm

The results presented in the previous section were obtained in the effective low-energy model neglecting the trigonal warping effects. This effective two-band model 
possesses complete rotational invariance \cite{McCann} in contrast with more restricted $C_3$ rotational invariance of the hexagonal lattice of BLG.
This more restricted symmetry is taken into account by the trigonal warping terms in the two-band model.
The corresponding low energy effective Hamiltonian is given by \cite{McCann}
\begin{align}
	H=\frac{v_F^2}{\gamma_1}\left(\begin{array}{cc}
		0 & \left(p_{-}\right)^2 \\
		\left(p_{+}\right)^2 & 0
	\end{array}\right)+\Delta\left(\begin{array}{cc}
		1 & 0 \\
		0 & -1
	\end{array}\right)- v_3\left(\begin{array}{cc}
		0 & p_+ \\
		p_- & 0
	\end{array}\right)+V(r),
\end{align}
where $v_3 \approx v_F/10$ accounts for the trigonal warping effects (without loss of generality we consider the valley $K$). The screened potential $V(r)$ of Coulomb impurity in the presence of gates is given by Eq.(\ref{potential-with-gates}) (in our calculations, we consider donors with unit charge $e$). Without the loss of generality, we
could consider energy dispersion in, e.g., the $K$ valley, which is given by
\begin{equation}
E=\sqrt{\frac{p^4}{4m^2}+v^2_3p^2-\frac{v_3p^3}{m}\cos(3\varphi)+\Delta^2}.
\label{trigonal-warping-dispersion}
\end{equation}
Clearly, this kinetic energy does not possess the complete rotational symmetry as terms with $v_3$ reduce this symmetry to the $C_3$ symmetry. The minimum of the kinetic 
energy $E_{\text{min}}=\Delta$ is realized at $\mathbf{p}=0$ and three nonzero values of momenta defined by $|\mathbf{p}|=2mv_3$ and $\varphi=2\pi n/3$ where $n=0,1,2$ so that $\cos(3\varphi)=1$.
The energy spectrum given by Eq.(\ref{trigonal-warping-dispersion}) is plotted in Fig.\ref{fig:trigonal-warping}(a) (schematic view) and its section at $p_y=0$ is shown by red dashed line in Fig.\ref{fig:trigonal-warping}(b). For comparison, the energy spectrum of the two-band model without trigonal warping corrections is plotted by red dashed line in Fig.\ref{fig:trigonal-warping}(b).

The energy dispersion (\ref{trigonal-warping-dispersion}) implies that the trigonal warping could be neglected for $(v_3/v_F)^2\gamma_1/2 < |E| < \gamma_1/4$ \cite{McCann} and is relevant at smaller energies. Numerically, the upper limit of validity of the parabolic spectrum is approximately
$W=\gamma_1/4 \approx 0.1$ \text{eV} and the lower one is $T=\frac{\gamma_1}{2}(v_3/v_F)^2 \approx 0.002$ \text{eV}. The dimensional ratio of the upper and lower limits is $W/T=50$.

Rescaling momenta by the inverse of the wave length $\lambda_1=\hbar v_F / \sqrt{\gamma_1 \Delta}$ and using the wave function in form (\ref{wavefunction}), we find
\begin{align}
	\label{eq:trigonal-wapring-1}
	& k^2 b_j(k)+a_j(k)-\frac{v_3 \sqrt{\gamma_1}}{v_F \sqrt{\Delta}} k b_{j-3}(k)-\xi \int_0^{Q} d p p a_j(p) V_{j-1}(k,p)=\epsilon a_j(k), \\
	\label{eq:trigonal-wapring-2}
	& k^2 a_j(k)-b_j(k)-\frac{v_3 \sqrt{\gamma_1}}{v_F \sqrt{\Delta}} k a_{j+3}(k)-\xi \int_0^{Q} d p p b_j(p) V_{j+1}(k,p)=\epsilon b_j(k).
\end{align}
where we use the double-screened potential (27) with $q = e$ (for simplicity, we neglect screening due to $\sigma$-bonds electrons).
This system of equations couples equations for components of the wave function with angular momentum $j$ to the wave function with $j+3$ and $j-3$ 
components. To determine solutions to such a system of equations, we should truncate this infinite chain of coupled equations at $j\pm 3n$ with integer $n \geq 1$. 
Coupling to other angular momentum components is taken to be zero truncating the system of equations. Let us consider as an example the case of the $j=1$ ground state. Then, in the case $n=1$, we take into account the nearest components and solve the system of equations for $j=1$, $j=4$, and $j=-2$ with kernel functions
$V_{0}(k,q),\,V_{2}(k,q)$, $V_{3}(k,q),\, V_{5}(k,q)$, and $V_{-3}(k,q),\, V_{-1}(k,q)$.

\begin{figure}
\centering
\includegraphics[scale=0.5]{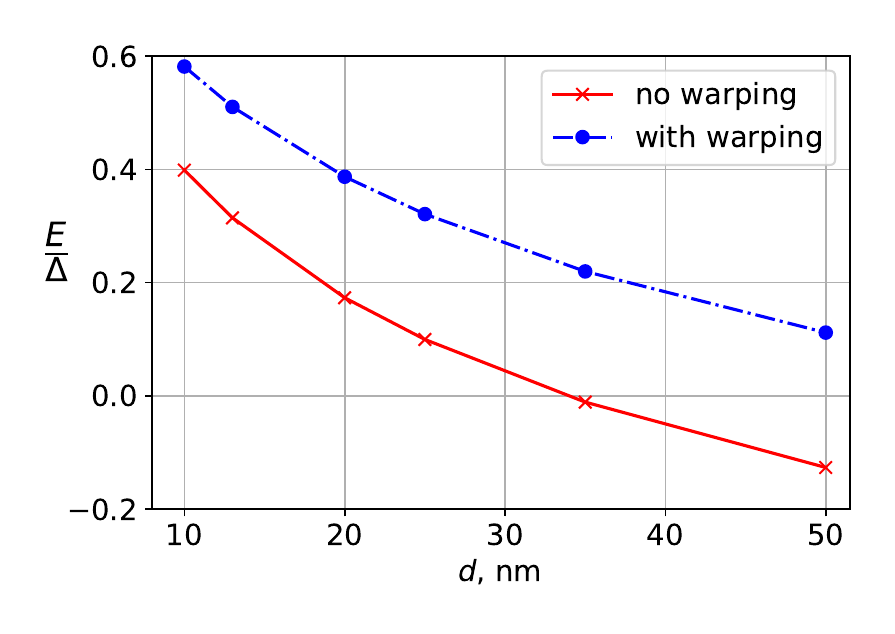} \includegraphics[scale=0.5]{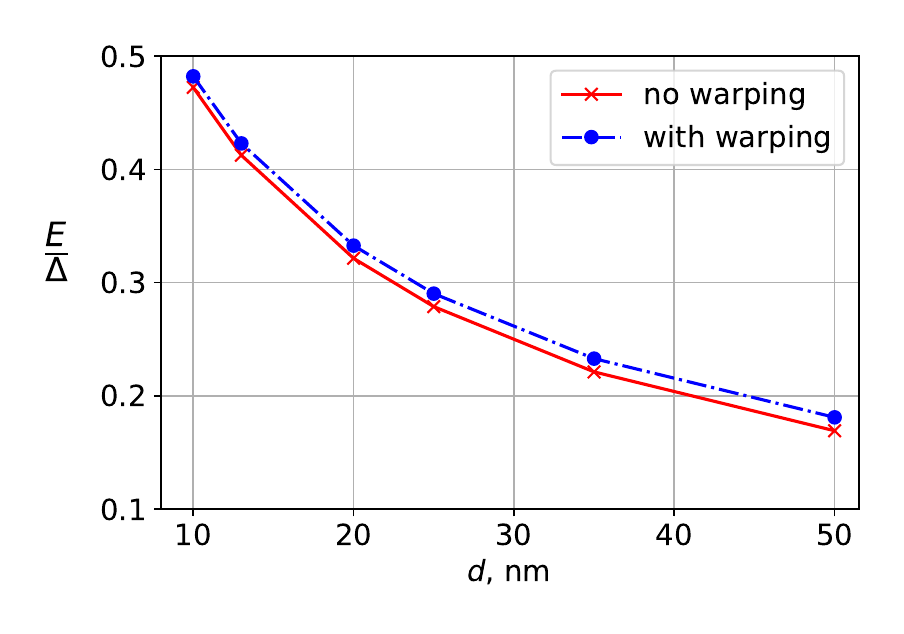}

	\caption{The donor ground state energy $E$ in units $\Delta$ as a function of distance to gates $d$ 
in the absence (red thick line) and with trigonal warping corrections (blue dash-dotted line) for $\Delta=1\, meV$ (left panel) and $\Delta=20 \,meV$ (right panel).
Here the cut-off parameter $n=1$ and the total angular momentum $j=1$. For all values of $d$, the trigonal warping effects result in somewhat higher energy of the electron bound state $E$ and hence smaller donor binding (ionization) energy $E_b = \Delta-E$.}
	\label{fig:trigonal-varping-results}
\end{figure}

The spectrum can be found by solving the system of equations \eqref{eq:trigonal-wapring-1} and \eqref{eq:trigonal-wapring-2}. For an angular momentum cut-off $n$,
the size of the discretized system is $2N(2n+1)$, where $N$ accounts for the discretization of momentum in a grid of $N$ points as was done in the previous section when
studying the electron bound states in the absence of the trigonal warping effects. To convert the wave function into a real space distribution, the Hankel transform of 
appropriate order should be used \cite{bilayer-instability}. In Fig.\ref{fig:trigonal-varping-results} we show the dependence of the lowest bound state energy with angular 
momentum $j=1$ in the presence of corrections from trigonal warping when components $j=-2$ and $j=4$ are taken into account. The effect of trigonal warping results in 
a small increase of the ground state energy for all values of gate distance. Corrections to the spectrum due to wave functions with higher components $j$, i.e., for $n \geq 2$, are approximately 10 times smaller compared to those with components $j=-2$ and $j=4$ for gap $\Delta\geq 1$ meV.


\section{Screened potential of charged impurity}
\label{sec:electric-potential}
\vspace{5mm}

Let us find the potential for the Coulomb impurity with charge $q$ situated outside BLG at $z^{\prime}$ with two
metallic gates at $z=d$ and $z=-d$ when BLG is at $z=0$. Without the loss of generality, we assume that the Coulomb
impurity is above BLG, i.e., $z^{\prime}>0$ and its $x$ and $y$ coordinates are $x=0$ and $y=0$. In addition, we assume also that the BLG sheet is separated from gates by a medium with dielectric constant $\kappa$. Then we have the following
Poisson equation for potential $\phi$ in region I ($0<z<d$):
\begin{equation}
\Delta \phi_I=-\frac{4\pi q}{\kappa}\delta(z-z^{\prime})\delta(x)\delta(y),
\label{Poisson-equation}
\end{equation}
where $\kappa \approx 4$ is the dielectric constant for hexagonal boron nitride. In addition, we have the boundary condition $\phi_I(x,y,z)=0$ at $z=d$. In region II ($-d<z<0$), the electric potential $\phi_{II}$ satisfies the Laplace equation
\begin{equation}
\Delta \phi_{II}=0
\label{Laplace-equation}
\end{equation}
with the boundary condition $\phi_{II}(x,y,z)=0$ at $z=-d$.

Note that the total electric potential is the superposition of two parts. One part describes constant electric field due to applied potentials $\pm U$ at the upper and lower gates, respectively (see, Fig.\ref{fig:intro-parameters}(a)). This part satisfies the Laplace equation. The other part satisfies the Poisson equation whose source is related to impurity charge and has trivial boundary conditions at gates $\phi_I(z=d)=\phi_{II}(z=-d)=0$. We determine this part in this section.

Electron quasiparticles in bilayer graphene can move when perturbed by an external electric field
resulting in screening of this field. Mathematically, this screening is defined by the induced charge density $\sigma(x,y)$ in BLG. This charge density contributes to the Poisson equation as follows:
\begin{equation}
\Delta \phi=-\frac{4\pi}{\kappa}\delta(z)\sigma(x,y).
\label{Poisson-equation-graphene}
\end{equation}
Integrating this equation over $z$ from $-\epsilon$ to $\epsilon$ and then setting $\epsilon \to 0$, we conclude that the induced charged density in BLG results in the following matching condition for the normal component of the electric field $E_z=-\partial_z \phi$:
\begin{equation}
E_z(0_+,x,y)=E_z(0_-,x,y)+\frac{4\pi}{\kappa}\sigma(x,y).
\label{normal-component-matching}
\end{equation}
Note that although the normal component of the electric field is discontinuous according to Eq.(\ref{normal-component-matching}) at $z=0$, potential $\phi(x,y,z)$ is well defined and continuous at $z=0$.

The charge density in bilayer graphene is defined via the Green function $G(u,u^{\prime})=-i\langle \psi(u)\psi^+(u^{\prime})\rangle$ as follows:
\begin{equation}
\sigma(x,y)=ie\,\mbox{tr}\,G(u,u^{\prime}),\quad u=(t,x,y),\quad u^{\prime}=(t^{\prime},x,y),\quad t^{\prime}-t=0_+,
\label{charge-density}
\end{equation}
where $-e<0$ is the electron charge. By making use of the Green function
$$
G(u,u^{\prime})=\langle u|\frac{\hbar}{i\hbar\partial_t-H+e\phi}|u^{\prime}\rangle,
$$
where $\phi$ is the potential at $z=0$, the induced charge density of the first order in $\phi$ is given by
\begin{equation}
\sigma(x,y)=-ie^2\int d^3u'\,\mbox{tr}\left[G_0(u-u^{\prime})\phi(u^{\prime})G_0(u^{\prime}-u)\right]=-e^2\int d^3u^{\prime}\,\Pi(u-u^{\prime})\phi(u^{\prime}),
\label{induced-charge-density}
\end{equation}
where $\Pi(u)=i\mbox{tr}\left[G_0(u)G_0(-u)\right]$ is the polarization function and $G_0(u)$ is the free Green function. Since we assume that $\phi(u^{\prime})$
does not depend on $t^{\prime}$, one can make the change of variable $t^{\prime}\to t^{\prime}-t$ in the integral in Eq.(\ref{induced-charge-density}). Therefore, the
induced charge density $\sigma(x,y)$ indeed does not depend on $t$.

Thus, Eq.(\ref{normal-component-matching}) gives the following matching conditions at plane $z=0$ where BLG is situated:
\begin{align}
\phi_I(\mathbf{x},z=0_+)=\phi_{II}(\mathbf{x},z=0_-),\quad \kappa\partial_z\phi_I(\mathbf{x},z=0_+)-\kappa\partial_z\phi_{II}(\mathbf{x},z=0_-)
=-4\pi\sigma(\mathbf{x}).
\end{align}
The induced surface charge density defined in Eq.(\ref{induced-charge-density}) equals
\begin{align}
\sigma(\mathbf{x})=-e^2\int dtd^2y\,\Pi(t,\mathbf{x}-\mathbf{y})\phi_I(\mathbf{y},0_+),\quad \mbox{or}\quad\sigma(\mathbf{x})=-e^2\int dtd^2y
\Pi(t,\mathbf{x}-\mathbf{y})\phi_{II}(\mathbf{y},0_-),
\end{align}
where we took into account that potentials are time independent.

It is convenient to perform the Fourier transform of the equations for the electric potential in regions I and II with respect to $x$ and $y$. Then we
obtain the equations
\begin{align}
(\partial_z^2-\mathbf{k}^2)\phi_I(\mathbf{k},z)=-\frac{4\pi q}{\kappa}\delta(z-z'),\quad (\partial_z^2-\mathbf{k}^2)\phi_{II}(\mathbf{k},z)=0,
\end{align}
whose general solutions have the form
\begin{align}
&\phi_I(\mathbf{k},z)=\frac{4\pi q}{\kappa}\frac{e^{-|\mathbf{k}||z-z'|}}{2|\mathbf{k}|}+A_1(\mathbf{k})e^{-|\mathbf{k}|z}+A_2(\mathbf{k})
e^{|\mathbf{k}|z},\\
&\phi_{II}(\mathbf{},z)=B_1(\mathbf{k})e^{-|\mathbf{k}| z}+B_2(\mathbf{k})e^{|\mathbf{k}|z},
\end{align}
where $A_1(\mathbf{k})$, $A_2(\mathbf{k})$, $B_1(\mathbf{k})$, $B_2(\mathbf{k})$ are arbitrary functions of wave vector $\mathbf{k}=(k_x,k_y)$. The boundary and matching conditions take the form
\begin{align}
&\phi_I(\mathbf{k},d)=0,\quad \phi_{II}(\mathbf{k},-d)=0,\quad \phi_I(\mathbf{k},0)=\phi_{II}(\mathbf{k},0),
\label{boundary-conditions-1}\\
& \partial_z\phi_I(\mathbf{k},0_+)-
\partial_z\phi_{II}(\mathbf{k},0_-)=-\frac{4\pi}{\kappa}\sigma(\mathbf{k})=\frac{4\pi e^2}{\kappa}\Pi(k_0=0,\mathbf{k})\phi_{II}(\mathbf{k},0),
\label{boundary-conditions-2}
\end{align}
where $\Pi(k_0=0,\mathbf{k})\equiv \Pi(\mathbf{k})$ is the static polarization. Equations (\ref{boundary-conditions-1}) and(\ref{boundary-conditions-2}) lead to a system of linear equations for
unknown functions $A_1,A_2,B_1,B_2$.
Solving this system we find the potential in the region $-d<z<0<z'<d$:
\begin{equation}
  \phi_{II}(k,z,z')=\frac{2\pi q}{\kappa}\frac{\sinh[k(d+z)]\sinh[k(d-z')]}{[k\cosh(kd)+\frac{2\pi e^2}{\kappa}\Pi(k)\sinh(kd)]\sinh(kd)}.
\label{potential-final-II}
\end{equation}
We note that this formula applies to any truly two-dimensional system with free charge carriers and polarization function $\Pi(k)$, for example, graphene.
In the absence of polarization, $\Pi(k)=0$, this potential reduces, after making the Fourier transform in $\mathbf{k}$, to the
potential of a point charge $q$ between parallel conducting plates (see Eq.(5) in Ref.\cite{Glasser}).

In the region $0<z<z'<d$ the potential has the form
\begin{equation}
  \phi_I(k,z,z')=\frac{2\pi q}{\kappa k}\frac{\sinh[k(d-z')][k\sinh[k(d+z)]+\frac{4\pi e^2}{\kappa}\Pi(k)\sinh(kd)\sinh(kz)]}
  {[k\cosh(kd)+\frac{2\pi e^2}{\kappa}\Pi(k)\sinh(kd)]\sinh(kd)}.
 \label{potential-final-I}
  \end{equation}

When gates are removed ($d\to\infty$) potential (\ref{potential-final-II}) reduces to
\begin{equation}
\phi_{II}(k)=\frac{2\pi q}{\kappa}\frac{e^{k(z-z')}}{k+\frac{2\pi e^2}{\kappa}\Pi(k)},
\label{general-formula-II}
\end{equation}
which for $z=z'=0$ in configuration space gives the interaction energy $-e\phi(r)$ of electron with charge $(-e)$ and impurity with charge
$q$ described by Eq.(27) in Ref.\cite{bilayer-instability}. For $z=z'=0$, potential (\ref{potential-final-II}) takes the form
\begin{equation}
  \phi(k)=\frac{2\pi q}{\kappa}\frac{1}{k\coth(kd)+\frac{2\pi e^2}{\kappa}\Pi(k)},
  \label{potential-with-gates}
\end{equation}
and, for the interaction energy, we obtain 
\begin{equation}
V(r)=-\frac{eq}{\kappa}\int\limits_0^\infty\frac{dk k J_0(k r)}{k\coth(kd)+\frac{2\pi e^2}{\kappa}\Pi(k) }.
\label{pot-energy}
\end{equation}

Note that this interaction energy does not include screening due to core electrons related to $\sigma$-bonds. It is known from studies in single \cite{Ulybyshev} and bilayer graphene \cite{Santos} that core electrons affect the polarization function. This contribution can be accounted by replacement $2\pi e^2\Pi(k)$ in Eq.(\ref{pot-energy}) with $2\pi e^2\Pi(k) + \sigma_b k$. In bilayer graphene $\sigma_b=1.8$ that gives dielectric constant  $\varepsilon \approx 2.8$ \cite{Santos}, which defines screening of the bare Coulomb potential in suspended bilayer graphene ($\kappa=1$) and in the absence of gates ($d \to \infty$) and vanishing polarization $\Pi(k)$ due to itinerant electrons.
We use the double screened potential (\ref{potential-with-gates}) and the interaction energy (\ref{pot-energy})
with single positive charge $q=e$
to determine the electron bound states defined by Eqs.(\ref{eq:system-potential-1}) and (\ref{eq:system-potential-2}) in Sec.\ref{sec:two-band-model} as well as Eqs.(\ref{eq:trigonal-wapring-1}) and (\ref{eq:trigonal-wapring-2}) in Sec.\ref{sec:trigonal-warping}. To gain insight into importance of screening due to core electrons, we analyze both cases $\sigma_b=0$ and $\sigma_b \ne 0$.

One can find more general formula when the upper gate is at distance $d_1$  and the lower gate is at other distance $d_2$ from graphene sheet
situiated at $z=0$. In addition, we assume that the BLG sheet is separated from gates by media with $\kappa_1$ (upper) and  $\kappa_2$ dielectric constants. The potential
produced by the charge $q$ sitting in \revisionA{the} upper medium  at the point $z'$ has the form ($0<z<z'<d_1$)
\begin{equation}
  \phi_I(k,z,z')=4\pi q\frac{\sinh[k(d_1-z')]\left[k\kappa_1\cosh(k z)\sinh(k d_2)+\left(k\kappa_2\cosh(k d_2)+4\pi e^2\Pi(k)\sinh(kd_2)
  \sinh(k z)\right)\right]}{k\kappa_1\left[k\kappa_2\cosh(kd_2)\sinh(kd_1)+\left(k\kappa_1\cosh(kd_1)+4\pi e^2\Pi(k)\sinh(kd_1)\right)
  \sinh(k d_2)\right]}
\end{equation}
It is easy to check that for $d_1=d_2=d$ and $\kappa_1=\kappa_2=\kappa$ this formula reduces to Eq.(\ref{potential-final-I}). If $z=z'=0$
we obtain the potential in graphene sheet
\begin{equation}
\phi(k)=\frac{4\pi q}{k\kappa_1\coth(kd_1)+k\kappa_2\coth(kd_2)+4\pi e^2\Pi(k)},
\end{equation}
which generalizes Eq.(\ref{potential-with-gates}) \cite{remark}. The last formula allows to study the case of one gate. For example, taking the limit
$d_2\to\infty$ we obtain
\begin{equation}
\phi(k)=\frac{4\pi q}{k\kappa_1\coth(kd_1)+k+4\pi e^2\Pi(k)}.
\label{potential-1gate}
\end{equation}

The polarization function in BLG can be approximated by the expression (see Eq.(30) in \cite{bilayer-instability})
\begin{equation}
\frac{2\pi e^2}{\kappa}\Pi(k)=\frac{k^2}{k_\Delta+\lambda_\gamma k^2},\quad k_\Delta=\frac{3\kappa}{4\alpha_g}\frac{1}{\lambda},
\quad \lambda_\gamma = \frac{\kappa\hbar v_F}{4\ln2\alpha_g\gamma_1}=\frac{\kappa}{4\ln2\alpha_g}a,
\label{polarization-approx}
\end{equation}
where $\lambda=\hbar v_F/\Delta$
and $a=\hbar v_F/\gamma_1 \approx 1.7\,\text{nm}$. 

For not too large values of $\kappa$, coefficients $k_\Delta$ and $\lambda_\gamma$ are of order $1/a$ and $a$, respectively. Thus, we have three scales: $a$, $d$, and $\lambda$ with $a\ll d\ll\lambda$ for BLG with gates. In fact, since we are 
not interested in distances less than $\lambda_\gamma\approx a$, we can set $\lambda_\gamma=0$. Then the potential takes the form
\begin{equation}
V(r)=-\frac{eq}{\kappa}\int\limits_0^\infty\frac{dk J_0(k r)}{\coth(kd)+k/k_\Delta}.
\label{potential-approx}
\end{equation}

For $r<d$, the main contribution to potential (\ref{potential-approx}) comes from the region with large $k$ such that $k d\gg1$ where
gate screening is not effective. Approximating in this region
$\coth(kd)\simeq1$, we get the integral (see Eq.(2.12.3.6) in \cite{PrudnikovII})
\begin{equation}
V(r)=-\frac{eq k_\Delta}{\kappa}\int\limits_0^\infty\frac{dk J_0(k r)}{k+k_\Delta}=-\frac{eq k_\Delta}{\kappa}\frac{\pi}{2}
\left[\mathbf{H}_0(k_\Delta r)-Y_0(k_\Delta r)\right],
\label{energy1-Struve}
\end{equation}
where $\mathbf{H}_0(x)$ and $Y_0(x)$ are the Struve function and the Bessel function of the second kind, respectively. This potential energy is known in the literature as the Rytova-Keldysh potential \cite{Rytova,Chaplik,Keldysh}. Thus, in the region
$r<d<\lambda$, the potential behaves as follows:
\begin{equation}
V(r)=-\frac{eq k_\Delta}{\kappa}\ln(2 e^{\gamma}/(rk_\Delta)),
\end{equation}
where $\gamma$ is the Euler–Mascheroni constant.

If we keep $\lambda_\gamma\neq0$ in (\ref{polarization-approx}), then we get more general function than function (\ref{energy1-Struve})
\begin{equation}
V(r)=-\frac{eq}{\kappa}\left[\frac{1}{r}-\frac{\pi}{2\lambda_\gamma(q_1-q_2)}\left\{q_1[\mathbf{H}_0(q_1r)-Y_0(q_1r)]-q_2
[\mathbf{H}_0(q_2r)-Y_0(q_2r)]\right\}\right],
\label{energy2-Struve}
\end{equation}
where $q_{1,2}=(1\pm\sqrt{1-4k_\Delta\lambda_\gamma})/2\lambda_\gamma$.
This expression was obtained earlier in \cite{bilayer-instability}. Similar mathematical expression (without $1/r$ term) for
the potential arises in the problem of the Coulomb interaction in a thin dielectric film in the presence of one gate \cite{Vinokur2017}.
Since $k_\Delta\lambda_\gamma\ll1$, we have $q_1\simeq1/\lambda_\gamma$, $k_\Delta\simeq k_\Delta$ and the interaction energy takes
the form
\begin{equation}
V(r)=-\frac{eq}{\kappa}\left[\frac{1}{r}-\frac{\pi}{2\lambda_\gamma}\left\{[\mathbf{H}_0(r/\lambda_\gamma)-Y_0(r/\lambda_\gamma)]-
k_\Delta\lambda_\gamma[\mathbf{H}_0(k_\Delta r)-Y_0(k_\Delta r)]\right\}\right].
\label{energy3-Struve}
\end{equation}

For $r>d$, the main contribution to the integral gives the region of small $k$ where we can replace $\coth x$ by its
two first terms of the Taylor expansion. Then we obtain
\begin{equation}
V(r)=-\frac{eqd}{\kappa}\int\limits_0^\infty\frac{dk k J_0(k r)}{1+b^2k^2}=-\frac{eqd}{\kappa b^2}K_0(r/b),\quad
b=d\left(1+\frac{4\alpha_g}{3\kappa}\frac{\lambda}{d}\right)^{1/2}.
\label{potential-K0}
\end{equation}
This potential is used in Appendix for an
analytical estimation of the electron binding energy.
Since $\lambda\gg d$, the asymptotic behavior is governed by length $b\approx\sqrt{d\lambda}$,
\begin{equation}
V(r)=-\frac{eqd}{\kappa b^2}\sqrt{\frac{\pi b}{2r}}e^{-r/b},\quad r>b.
\end{equation}
For $r<b$, we have
\begin{equation}
V(r)=-\frac{eqd}{\kappa b^2}\ln\frac{2e^{-\gamma}b}{r}.
\end{equation}
Thus, the logarithmic behavior persists up to distances $\sim\sqrt{d\lambda}$. We note that in the case of one gate we would have a linear in $k$
term in the denominator of the integrand in Eq.(\ref{potential-K0}) and this changes the asymptotic behavior at large distance from the exponential to power law $\sim 1/r^3$ \cite{Vinokur2017}.

\section{Conclusions}
\label{sec:conclusions}

\vspace{5mm}

In modern devices bilayer graphene is enveloped by two insulating layers of hexagonal boron nitride and has top and bottom metallic gates made of graphite or gold, for example, at the same distance $d$ from BLG. These gates play twofold role for charged impurities. First, the electric field created by potential difference between top and bottom gates produces a gap $2\Delta$ in the electron spectrum of BLG. As a result, donors and acceptors located in BLG create electron and hole bound states in the gap. Second, gates screen the Coulomb potential of charged impurities in BLG. We studied interplay of this screening with BLG polarization screening. We derived general formulas given by Eqs.(\ref{potential-final-I})-(\ref{potential-with-gates}) for the double screened electric potential of a charged impurity. Using this double screened electric potential of a singly charged donor we calculated its binding (ionization) energy $E_b$ solving numerically the integral equation for electron bound states. 
We analyzed also the role of screening due to $\sigma$-bonds electrons and found it to be rather minor. We checked our numerical results via a simple analytical estimate of the electron binding energy of donor using the uncertainty principle.

We found that $E_b$ is close to the half of the gap. Namely, our results show that for distances to gates $10 < d < 200$\,\text{nm} and gaps $1 < \Delta < 70$\,\text{meV} the ratio $E_b/\Delta$ changes from 0.4 to 1.4. The ratio $E_b/\Delta$ stays close to unity because of the dominating role of the BLG polarization screening which at any $\Delta$ reduces the depth of the Coulomb potential well to $\sim \Delta$. 
We also studied the effect of the trigonal warping of BLG energy bands and found that even at very small $\Delta = 1\,\text{meV}$ it leads only to 20 percent reduction of ratio $E_b/\Delta$. At much larger $\Delta = 20\,\text{meV}$  the trigonal warping effects practically play no role for $E_b/\Delta$. Of course, our results for the electron binding energy to donor are applicable also to the binding energy of a hole and a negative acceptor.

The Coulomb potential of a donor located in hBN calculated in Sec. IV can be used for evaluation of the binding energy of such a donor in future studies. Assuming that donors are randomly distributed in hBN one can calculate the density of states of BLG electrons bound to hBN donors, which in turn can be used for the calculation of hopping conductivity of BLG. While this paper dealt with relatively small gaps $\Delta < 70$\,\text{meV} employing the two band model, the study of both BLG and hBN donors can be extended to $\Delta > 70$\,\text{meV} with the use of the BLG four band model. Finally, it would be interesting to study BLG donors in the presence of an external magnetic field. We would like to remind that the spectrum of a Coulomb impurity in bilayer graphene in a magnetic field was previously investigated in Ref.\cite{Peeters} but in the absence of double screening due to the BLG polarization and gates.

\section*{Acknowledgments}
The work of E.V.G. and V.P.G. was supported by the Program "Dynamics of particles and collective excitations in high-energy physics, astrophysics and quantum macrosystems" of the Department of Physics and Astronomy of the NAS of Ukraine. V.P.G. thanks the Simons Foundation for the partial financial support. D.O.O. acknowledges the support from the Netherlands Organization for Scientific Research (NWO/OCW) and from the European Research Council (ERC) under the European Union's Horizon 2020 research and innovation program.
\vspace{5mm}

\appendix
\section{Uncertainty principle estimate of electron binding energy of donor}
\label{sec:appendix-A}

\vskip5mm

To gain an analytic insight into the obtained numerical results, we consider in this appendix an approximate method of derivation of the donor binding energy  which uses the uncertainty principle. Although the accuracy of this method is not under control, its obvious advantage is simplicity and transparency.

To apply the uncertainty principle, we begin with the expression for the classical energy of the electron in the upper band
\begin{equation}
E(r)=\sqrt{({\bf p^2(\mathbf{r})}/2m)^2+\Delta^2}+V(r),
\label{quasiclassical-energy}
\end{equation}
where $V$ is the potential energy for which we use the approximate analytic 
potential (\ref{potential-K0}). Since the potential energy does not depend on angle $\phi$, it is natural to expect that the wave function of the lowest energy 
state does not depend on angle also, i.e., it is a function of radial coordinate $r$ only. Consequently, we should apply the uncertainty principle to the radial
momentum.

In the corresponding quantum-mechanical problem, the classical momentum $\mathbf{p}^2(\mathbf{r})$ in Eq.(\ref{quasiclassical-energy})
is replaced by the Laplace operator $\mathbf{p}^2=-\hbar^2\Delta$.
In 2D, the radial part of the Laplace operator equals $\Delta_r=\partial^2_r+1/r\,\partial_r$.  On the other hand, the Hermitian radial momentum is given by 
$\hat{p}_r=-i\hbar(\partial_r+1/(2r))$ in view of the presence of the Jacobian $r$ in the integral over polar coordinates (for a similar definition of the Hermitian radial momentum operator in 3D, see Ref.\cite{Flugge}). Then it is easy to check that $-\hbar^2\Delta_r=\hat{p}^2_r-\hbar^2/(4r^2)$. This means that the correct application of the 
uncertainty relation for the radial momentum and radial coordinate in 2D should be made via the replacement
$-\hbar^2\Delta_r=\hat{p}^2_r-\hbar^2/(4r^2) \to \hbar^2/r^2-\hbar^2/(4r^2)=3\hbar^2/(4r^2)$.

\begin{figure}[ht]
\includegraphics[width=.4\textwidth]{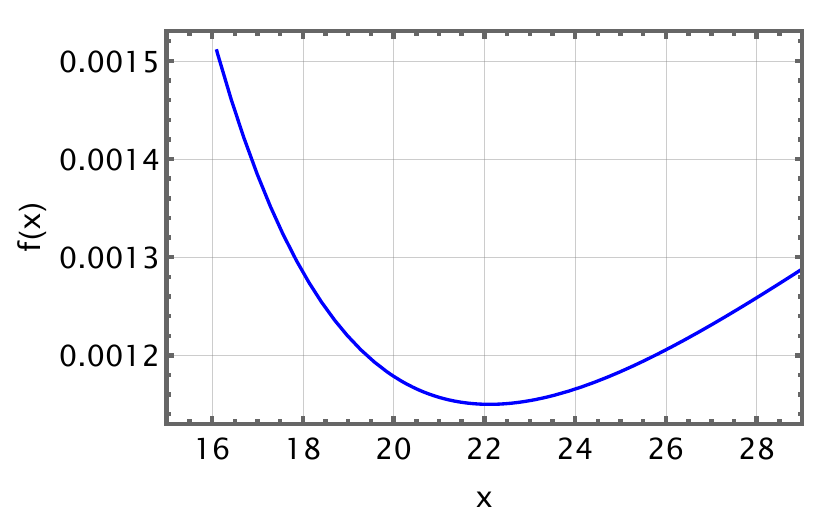}
\caption{The average energy (\ref{average-energy-screened-correct}) in the two-band model in units $\gamma_1$ as a function of distance $x=r/a$ for $\Delta=1$ meV
($\delta=0.00256$)
and distance $d=13$ nm to gates.}
\label{2band-energy-K0-correct}
\end{figure}

Applying this prescription to function (\ref{quasiclassical-energy}), we arrive at the problem to determine a minimum of the function
\begin{equation}
f(x)=\frac{E(r)}{\gamma_1}=\sqrt{\frac{9}{16x^4}+\delta^2}-\frac{\alpha_g}{\kappa}\frac{a}{d+\frac{4\alpha_g}{3\kappa}\frac{a}{\delta}}
K_0\left(x\frac{a}{d(1+\frac{4\alpha_g}{3\kappa}\frac{a}{d\delta})^{1/2}}\right),
\label{average-energy-screened-correct}
\end{equation}
where $\delta=\Delta/\gamma_1$ and $x=r/a$ is dimensionless distance with $a=\hbar v_F/\gamma_1 \approx 1.7\,\text{nm}$. This function is plotted in Fig.\ref{2band-energy-K0-correct} for
$d=13\,\text{nm}$,
$\Delta=1\,\text{meV}$ and has minimum at $x=22.11$ which is equal to $0.0011$ that gives $E=0.43\,\text{meV}$ in dimensional units.
This value should be compared with $E=0.31\,\text{meV}$ in Table I obtained in numerical calculations. As usual, a variational method gives an overestimated value for the bound state energy compared to the exact computation. Still, as one can see, it gives a resonably good estimate of the electron bound state energy.

\vspace{5mm}

We would like to note also that the interaction energy (\ref{potential-approx}) starts to deviate from the analytical expression (\ref{potential-K0}) \revisionA{for} $x<4$.
However, in this region, the kinetic energy significantly exceeds the potential energy and the minimum of total energy lies in the region $x\gg4$,
thus, the calculation with potential (\ref{potential-K0}) is valid.
\vspace{5mm}

\bibliography{bibliography_bilayer}

\begin{thebibliography}{30}%
\makeatletter
\providecommand \@ifxundefined [1]{%
 \@ifx{#1\undefined}
}%
\providecommand \@ifnum [1]{%
 \ifnum #1\expandafter \@firstoftwo
 \else \expandafter \@secondoftwo
 \fi
}%
\providecommand \@ifx [1]{%
 \ifx #1\expandafter \@firstoftwo
 \else \expandafter \@secondoftwo
 \fi
}%
\providecommand \natexlab [1]{#1}%
\providecommand \enquote  [1]{``#1''}%
\providecommand \bibnamefont  [1]{#1}%
\providecommand \bibfnamefont [1]{#1}%
\providecommand \citenamefont [1]{#1}%
\providecommand \href@noop [0]{\@secondoftwo}%
\providecommand \href [0]{\begingroup \@sanitize@url \@href}%
\providecommand \@href[1]{\@@startlink{#1}\@@href}%
\providecommand \@@href[1]{\endgroup#1\@@endlink}%
\providecommand \@sanitize@url [0]{\catcode `\\12\catcode `\$12\catcode
  `\&12\catcode `\#12\catcode `\^12\catcode `\_12\catcode `\%12\relax}%
\providecommand \@@startlink[1]{}%
\providecommand \@@endlink[0]{}%
\providecommand \url  [0]{\begingroup\@sanitize@url \@url }%
\providecommand \@url [1]{\endgroup\@href {#1}{\urlprefix }}%
\providecommand \urlprefix  [0]{URL }%
\providecommand \Eprint [0]{\href }%
\providecommand \doibase [0]{https://doi.org/}%
\providecommand \selectlanguage [0]{\@gobble}%
\providecommand \bibinfo  [0]{\@secondoftwo}%
\providecommand \bibfield  [0]{\@secondoftwo}%
\providecommand \translation [1]{[#1]}%
\providecommand \BibitemOpen [0]{}%
\providecommand \bibitemStop [0]{}%
\providecommand \bibitemNoStop [0]{.\EOS\space}%
\providecommand \EOS [0]{\spacefactor3000\relax}%
\providecommand \BibitemShut  [1]{\csname bibitem#1\endcsname}%
\let\auto@bib@innerbib\@empty
\bibitem [{\citenamefont {McCann}\ and\ \citenamefont
  {Fal’ko}(2006)}]{McCann}%
  \BibitemOpen
  \bibfield  {author} {\bibinfo {author} {\bibfnamefont {E.}~\bibnamefont
  {McCann}}\ and\ \bibinfo {author} {\bibfnamefont {V.~I.}\ \bibnamefont
  {Fal’ko}},\ }\bibfield  {title} {\bibinfo {title} {{L}andau-level
  degeneracy and quantum {H}all effect in a graphite bilayer},\ }\href
  {https://doi.org/10.1103/physrevlett.96.086805} {\bibfield  {journal}
  {\bibinfo  {journal} {Phys. Rev. Lett.}\ }\textbf {\bibinfo {volume} {96}},\
  \bibinfo {pages} {086805} (\bibinfo {year} {2006})}\BibitemShut {NoStop}%
\bibitem [{\citenamefont {Novoselov}\ \emph {et~al.}(2006)\citenamefont
  {Novoselov}, \citenamefont {McCann}, \citenamefont {Morozov}, \citenamefont
  {Fal’ko}, \citenamefont {Katsnelson}, \citenamefont {Zeitler},
  \citenamefont {Jiang}, \citenamefont {Schedin},\ and\ \citenamefont
  {Geim}}]{Novoselov}%
  \BibitemOpen
  \bibfield  {author} {\bibinfo {author} {\bibfnamefont {K.~S.}\ \bibnamefont
  {Novoselov}}, \bibinfo {author} {\bibfnamefont {E.}~\bibnamefont {McCann}},
  \bibinfo {author} {\bibfnamefont {S.~V.}\ \bibnamefont {Morozov}}, \bibinfo
  {author} {\bibfnamefont {V.~I.}\ \bibnamefont {Fal’ko}}, \bibinfo {author}
  {\bibfnamefont {M.~I.}\ \bibnamefont {Katsnelson}}, \bibinfo {author}
  {\bibfnamefont {U.}~\bibnamefont {Zeitler}}, \bibinfo {author} {\bibfnamefont
  {D.}~\bibnamefont {Jiang}}, \bibinfo {author} {\bibfnamefont
  {F.}~\bibnamefont {Schedin}},\ and\ \bibinfo {author} {\bibfnamefont {A.~K.}\
  \bibnamefont {Geim}},\ }\bibfield  {title} {\bibinfo {title} {Unconventional
  quantum {H}all effect and {B}erry\'{}s phase of $2\pi$ in bilayer graphene},\
  }\href {https://doi.org/10.1038/nphys245} {\bibfield  {journal} {\bibinfo
  {journal} {Nature Phys.}\ }\textbf {\bibinfo {volume} {2}},\ \bibinfo {pages}
  {177} (\bibinfo {year} {2006})}\BibitemShut {NoStop}%
\bibitem [{\citenamefont {Castro~Neto}\ \emph {et~al.}(2009)\citenamefont
  {Castro~Neto}, \citenamefont {Guinea}, \citenamefont {Peres}, \citenamefont
  {Novoselov},\ and\ \citenamefont {Geim}}]{Neto}%
  \BibitemOpen
  \bibfield  {author} {\bibinfo {author} {\bibfnamefont {A.~H.}\ \bibnamefont
  {Castro~Neto}}, \bibinfo {author} {\bibfnamefont {F.}~\bibnamefont {Guinea}},
  \bibinfo {author} {\bibfnamefont {N.~M.~R.}\ \bibnamefont {Peres}}, \bibinfo
  {author} {\bibfnamefont {K.~S.}\ \bibnamefont {Novoselov}},\ and\ \bibinfo
  {author} {\bibfnamefont {A.~K.}\ \bibnamefont {Geim}},\ }\bibfield  {title}
  {\bibinfo {title} {The electronic properties of graphene},\ }\href
  {https://doi.org/10.1103/revmodphys.81.109} {\bibfield  {journal} {\bibinfo
  {journal} {Rev. Mod. Phys.}\ }\textbf {\bibinfo {volume} {81}},\ \bibinfo
  {pages} {109} (\bibinfo {year} {2009})}\BibitemShut {NoStop}%
\bibitem [{\citenamefont {Castro}\ \emph {et~al.}(2007)\citenamefont {Castro},
  \citenamefont {Novoselov}, \citenamefont {Morozov}, \citenamefont {Peres},
  \citenamefont {dos Santos}, \citenamefont {Nilsson}, \citenamefont {Guinea},
  \citenamefont {Geim},\ and\ \citenamefont {Neto}}]{Castro}%
  \BibitemOpen
  \bibfield  {author} {\bibinfo {author} {\bibfnamefont {E.~V.}\ \bibnamefont
  {Castro}}, \bibinfo {author} {\bibfnamefont {K.~S.}\ \bibnamefont
  {Novoselov}}, \bibinfo {author} {\bibfnamefont {S.~V.}\ \bibnamefont
  {Morozov}}, \bibinfo {author} {\bibfnamefont {N.~M.~R.}\ \bibnamefont
  {Peres}}, \bibinfo {author} {\bibfnamefont {J.~M. B.~L.}\ \bibnamefont {dos
  Santos}}, \bibinfo {author} {\bibfnamefont {J.}~\bibnamefont {Nilsson}},
  \bibinfo {author} {\bibfnamefont {F.}~\bibnamefont {Guinea}}, \bibinfo
  {author} {\bibfnamefont {A.~K.}\ \bibnamefont {Geim}},\ and\ \bibinfo
  {author} {\bibfnamefont {A.~H.~C.}\ \bibnamefont {Neto}},\ }\bibfield
  {title} {\bibinfo {title} {Biased bilayer graphene: Semiconductor with a gap
  tunable by the electric field effect},\ }\href
  {https://doi.org/10.1103/physrevlett.99.216802} {\bibfield  {journal}
  {\bibinfo  {journal} {Phys. Rev. Lett.}\ }\textbf {\bibinfo {volume} {99}},\
  \bibinfo {pages} {216802} (\bibinfo {year} {2007})}\BibitemShut {NoStop}%
\bibitem [{\citenamefont {Zibrov}\ \emph {et~al.}(2017)\citenamefont {Zibrov},
  \citenamefont {Kometter}, \citenamefont {Zhou}, \citenamefont {Spanton},
  \citenamefont {Taniguchi}, \citenamefont {Watanabe}, \citenamefont
  {Zaletel},\ and\ \citenamefont {Young}}]{Young2017}%
  \BibitemOpen
  \bibfield  {author} {\bibinfo {author} {\bibfnamefont {A.~A.}\ \bibnamefont
  {Zibrov}}, \bibinfo {author} {\bibfnamefont {C.}~\bibnamefont {Kometter}},
  \bibinfo {author} {\bibfnamefont {H.}~\bibnamefont {Zhou}}, \bibinfo {author}
  {\bibfnamefont {E.~M.}\ \bibnamefont {Spanton}}, \bibinfo {author}
  {\bibfnamefont {T.}~\bibnamefont {Taniguchi}}, \bibinfo {author}
  {\bibfnamefont {K.}~\bibnamefont {Watanabe}}, \bibinfo {author}
  {\bibfnamefont {M.~P.}\ \bibnamefont {Zaletel}},\ and\ \bibinfo {author}
  {\bibfnamefont {A.~F.}\ \bibnamefont {Young}},\ }\bibfield  {title} {\bibinfo
  {title} {Tunable interacting composite fermion phases in a half-filled
  bilayer-graphene {L}andau level},\ }\href
  {https://doi.org/10.1038/nature23893} {\bibfield  {journal} {\bibinfo
  {journal} {Nature}\ }\textbf {\bibinfo {volume} {549}},\ \bibinfo {pages}
  {360} (\bibinfo {year} {2017})}\BibitemShut {NoStop}%
\bibitem [{\citenamefont {Joucken}\ \emph {et~al.}(2021)\citenamefont
  {Joucken}, \citenamefont {Bena}, \citenamefont {Ge}, \citenamefont
  {Quezada-Lopez}, \citenamefont {Pinon}, \citenamefont {Kaladzhyan},
  \citenamefont {Taniguchi}, \citenamefont {Watanabe}, \citenamefont
  {Ferreira},\ and\ \citenamefont {Velasco}}]{Joucken2021}%
  \BibitemOpen
  \bibfield  {author} {\bibinfo {author} {\bibfnamefont {F.}~\bibnamefont
  {Joucken}}, \bibinfo {author} {\bibfnamefont {C.}~\bibnamefont {Bena}},
  \bibinfo {author} {\bibfnamefont {Z.}~\bibnamefont {Ge}}, \bibinfo {author}
  {\bibfnamefont {E.}~\bibnamefont {Quezada-Lopez}}, \bibinfo {author}
  {\bibfnamefont {S.}~\bibnamefont {Pinon}}, \bibinfo {author} {\bibfnamefont
  {V.}~\bibnamefont {Kaladzhyan}}, \bibinfo {author} {\bibfnamefont
  {T.}~\bibnamefont {Taniguchi}}, \bibinfo {author} {\bibfnamefont
  {K.}~\bibnamefont {Watanabe}}, \bibinfo {author} {\bibfnamefont
  {A.}~\bibnamefont {Ferreira}},\ and\ \bibinfo {author} {\bibfnamefont
  {J.}~\bibnamefont {Velasco}},\ }\bibfield  {title} {\bibinfo {title} {Direct
  visualization of native defects in graphite and their effect on the
  electronic properties of bernal-stacked bilayer graphene},\ }\href
  {https://doi.org/10.1021/acs.nanolett.1c01442} {\bibfield  {journal}
  {\bibinfo  {journal} {Nano Lett.}\ }\textbf {\bibinfo {volume} {21}},\
  \bibinfo {pages} {7100} (\bibinfo {year} {2021})}\BibitemShut {NoStop}%
\bibitem [{\citenamefont {Zhou}\ \emph {et~al.}(2022)\citenamefont {Zhou},
  \citenamefont {Holleis}, \citenamefont {Saito}, \citenamefont {Cohen},
  \citenamefont {Huynh}, \citenamefont {Patterson}, \citenamefont {Yang},
  \citenamefont {Taniguchi}, \citenamefont {Watanabe},\ and\ \citenamefont
  {Young}}]{Young2022}%
  \BibitemOpen
  \bibfield  {author} {\bibinfo {author} {\bibfnamefont {H.}~\bibnamefont
  {Zhou}}, \bibinfo {author} {\bibfnamefont {L.}~\bibnamefont {Holleis}},
  \bibinfo {author} {\bibfnamefont {Y.}~\bibnamefont {Saito}}, \bibinfo
  {author} {\bibfnamefont {L.}~\bibnamefont {Cohen}}, \bibinfo {author}
  {\bibfnamefont {W.}~\bibnamefont {Huynh}}, \bibinfo {author} {\bibfnamefont
  {C.~L.}\ \bibnamefont {Patterson}}, \bibinfo {author} {\bibfnamefont
  {F.}~\bibnamefont {Yang}}, \bibinfo {author} {\bibfnamefont {T.}~\bibnamefont
  {Taniguchi}}, \bibinfo {author} {\bibfnamefont {K.}~\bibnamefont
  {Watanabe}},\ and\ \bibinfo {author} {\bibfnamefont {A.~F.}\ \bibnamefont
  {Young}},\ }\bibfield  {title} {\bibinfo {title} {Isospin magnetism and
  spin-polarized superconductivity in {B}ernal bilayer graphene},\ }\href
  {https://doi.org/10.1126/science.abm8386} {\bibfield  {journal} {\bibinfo
  {journal} {Science}\ }\textbf {\bibinfo {volume} {375}},\ \bibinfo {pages}
  {774} (\bibinfo {year} {2022})}\BibitemShut {NoStop}%
\bibitem [{\citenamefont {Silvestrov}\ and\ \citenamefont
  {Recher}(2017)}]{Silvestrov}%
  \BibitemOpen
  \bibfield  {author} {\bibinfo {author} {\bibfnamefont {P.~G.}\ \bibnamefont
  {Silvestrov}}\ and\ \bibinfo {author} {\bibfnamefont {P.}~\bibnamefont
  {Recher}},\ }\bibfield  {title} {\bibinfo {title} {Wigner crystal phases in
  bilayer graphene},\ }\href {https://doi.org/10.1103/physrevb.95.075438}
  {\bibfield  {journal} {\bibinfo  {journal} {Phys. Rev. B}\ }\textbf {\bibinfo
  {volume} {95}},\ \bibinfo {pages} {075438} (\bibinfo {year}
  {2017})}\BibitemShut {NoStop}%
\bibitem [{\citenamefont {Oriekhov}\ \emph {et~al.}(2017)\citenamefont
  {Oriekhov}, \citenamefont {Sobol}, \citenamefont {Gorbar},\ and\
  \citenamefont {Gusynin}}]{bilayer-instability}%
  \BibitemOpen
  \bibfield  {author} {\bibinfo {author} {\bibfnamefont {D.~O.}\ \bibnamefont
  {Oriekhov}}, \bibinfo {author} {\bibfnamefont {O.~O.}\ \bibnamefont {Sobol}},
  \bibinfo {author} {\bibfnamefont {E.~V.}\ \bibnamefont {Gorbar}},\ and\
  \bibinfo {author} {\bibfnamefont {V.~P.}\ \bibnamefont {Gusynin}},\
  }\bibfield  {title} {\bibinfo {title} {Coulomb center instability in bilayer
  graphene},\ }\href {https://doi.org/10.1103/physrevb.96.165403} {\bibfield
  {journal} {\bibinfo  {journal} {Phys. Rev. B}\ }\textbf {\bibinfo {volume}
  {96}},\ \bibinfo {pages} {165403} (\bibinfo {year} {2017})}\BibitemShut
  {NoStop}%
\bibitem [{\citenamefont {Ulybyshev}\ \emph {et~al.}(2013)\citenamefont
  {Ulybyshev}, \citenamefont {Buividovich}, \citenamefont {Katsnelson},\ and\
  \citenamefont {Polikarpov}}]{Ulybyshev}%
  \BibitemOpen
  \bibfield  {author} {\bibinfo {author} {\bibfnamefont {M.~V.}\ \bibnamefont
  {Ulybyshev}}, \bibinfo {author} {\bibfnamefont {P.~V.}\ \bibnamefont
  {Buividovich}}, \bibinfo {author} {\bibfnamefont {M.~I.}\ \bibnamefont
  {Katsnelson}},\ and\ \bibinfo {author} {\bibfnamefont {M.~I.}\ \bibnamefont
  {Polikarpov}},\ }\bibfield  {title} {\bibinfo {title} {Monte {C}arlo study
  of the semimetal-insulator phase transition in monolayer graphene with a
  realistic interelectron interaction potential},\ }\href
  {https://doi.org/10.1103/physrevlett.111.056801} {\bibfield  {journal}
  {\bibinfo  {journal} {Physical Review Letters}\ }\textbf {\bibinfo {volume}
  {111}},\ \bibinfo {pages} {056801} (\bibinfo {year} {2013})}\BibitemShut
  {NoStop}%
\bibitem [{\citenamefont {Santos}\ and\ \citenamefont
  {Kaxiras}(2013)}]{Santos}%
  \BibitemOpen
  \bibfield  {author} {\bibinfo {author} {\bibfnamefont {E.~J.~G.}\
  \bibnamefont {Santos}}\ and\ \bibinfo {author} {\bibfnamefont
  {E.}~\bibnamefont {Kaxiras}},\ }\bibfield  {title} {\bibinfo {title}
  {Electric-field dependence of the effective dielectric constant in
  graphene},\ }\href {https://doi.org/10.1021/nl303611v} {\bibfield  {journal}
  {\bibinfo  {journal} {Nano Letters}\ }\textbf {\bibinfo {volume} {13}},\
  \bibinfo {pages} {898} (\bibinfo {year} {2013})}\BibitemShut {NoStop}%
\bibitem [{\citenamefont {Gamayun}(2011)}]{Gamayun2011PRB}%
  \BibitemOpen
  \bibfield  {author} {\bibinfo {author} {\bibfnamefont {O.~V.}\ \bibnamefont
  {Gamayun}},\ }\bibfield  {title} {\bibinfo {title} {Dynamical screening in
  bilayer graphene},\ }\href {https://doi.org/10.1103/physrevb.84.085112}
  {\bibfield  {journal} {\bibinfo  {journal} {Phys. Rev. B}\ }\textbf {\bibinfo
  {volume} {84}},\ \bibinfo {pages} {085112} (\bibinfo {year}
  {2011})}\BibitemShut {NoStop}%
\bibitem [{\citenamefont {Nandkishore}\ and\ \citenamefont
  {Levitov}(2010{\natexlab{a}})}]{Nandkishore2010PRL_exciton}%
  \BibitemOpen
  \bibfield  {author} {\bibinfo {author} {\bibfnamefont {R.}~\bibnamefont
  {Nandkishore}}\ and\ \bibinfo {author} {\bibfnamefont {L.}~\bibnamefont
  {Levitov}},\ }\bibfield  {title} {\bibinfo {title} {Dynamical screening and
  excitonic instability in bilayer graphene},\ }\href
  {https://doi.org/10.1103/physrevlett.104.156803} {\bibfield  {journal}
  {\bibinfo  {journal} {Phys. Rev. Lett.}\ }\textbf {\bibinfo {volume} {104}},\
  \bibinfo {pages} {156803} (\bibinfo {year} {2010}{\natexlab{a}})}\BibitemShut
  {NoStop}%
\bibitem [{\citenamefont {Nandkishore}\ and\ \citenamefont
  {Levitov}(2010{\natexlab{b}})}]{Nandkishore2010PRB}%
  \BibitemOpen
  \bibfield  {author} {\bibinfo {author} {\bibfnamefont {R.}~\bibnamefont
  {Nandkishore}}\ and\ \bibinfo {author} {\bibfnamefont {L.}~\bibnamefont
  {Levitov}},\ }\bibfield  {title} {\bibinfo {title} {Electron interactions in
  bilayer graphene: {M}arginal {F}ermi liquid and zero-bias anomaly},\ }\href
  {https://doi.org/10.1103/physrevb.82.115431} {\bibfield  {journal} {\bibinfo
  {journal} {Physical Review B}\ }\textbf {\bibinfo {volume} {82}},\ \bibinfo
  {pages} {115431} (\bibinfo {year} {2010}{\natexlab{b}})}\BibitemShut
  {NoStop}%
\bibitem [{\citenamefont {Nandkishore}\ and\ \citenamefont
  {Levitov}(2010{\natexlab{c}})}]{Nandkishore2010PRB2}%
  \BibitemOpen
  \bibfield  {author} {\bibinfo {author} {\bibfnamefont {R.}~\bibnamefont
  {Nandkishore}}\ and\ \bibinfo {author} {\bibfnamefont {L.}~\bibnamefont
  {Levitov}},\ }\bibfield  {title} {\bibinfo {title} {Quantum anomalous {H}all
  state in bilayer graphene},\ }\href
  {https://doi.org/10.1103/physrevb.82.115124} {\bibfield  {journal} {\bibinfo
  {journal} {Phys. Rev. B}\ }\textbf {\bibinfo {volume} {82}},\ \bibinfo
  {pages} {115124} (\bibinfo {year} {2010}{\natexlab{c}})}\BibitemShut
  {NoStop}%
\bibitem [{\citenamefont {Pumplin}(1969)}]{Pumplin}%
  \BibitemOpen
  \bibfield  {author} {\bibinfo {author} {\bibfnamefont {J.}~\bibnamefont
  {Pumplin}},\ }\bibfield  {title} {\bibinfo {title} {Application of
  {S}ommerfeld-{W}atson transformation to an electrostatics problem},\ }\href
  {https://doi.org/10.1119/1.1975793} {\bibfield  {journal} {\bibinfo
  {journal} {Am. J. Phys.}\ }\textbf {\bibinfo {volume} {37}},\ \bibinfo
  {pages} {737} (\bibinfo {year} {1969})}\BibitemShut {NoStop}%
\bibitem [{\citenamefont {Glasser}(1970)}]{Glasser}%
  \BibitemOpen
  \bibfield  {author} {\bibinfo {author} {\bibfnamefont {M.~L.}\ \bibnamefont
  {Glasser}},\ }\bibfield  {title} {\bibinfo {title} {The potential of a point
  charge between capacitor plates},\ }\href {https://doi.org/10.1119/1.1976356}
  {\bibfield  {journal} {\bibinfo  {journal} {Am. J. Phys.}\ }\textbf {\bibinfo
  {volume} {38}},\ \bibinfo {pages} {415} (\bibinfo {year} {1970})}\BibitemShut
  {NoStop}%
\bibitem [{\citenamefont {McCann}\ \emph {et~al.}(2007)\citenamefont {McCann},
  \citenamefont {Abergel},\ and\ \citenamefont {Fal’ko}}]{McCann2007gamma1}%
  \BibitemOpen
  \bibfield  {author} {\bibinfo {author} {\bibfnamefont {E.}~\bibnamefont
  {McCann}}, \bibinfo {author} {\bibfnamefont {D.~S.~L.}\ \bibnamefont
  {Abergel}},\ and\ \bibinfo {author} {\bibfnamefont {V.~I.}\ \bibnamefont
  {Fal’ko}},\ }\bibfield  {title} {\bibinfo {title} {The low energy
  electronic band structure of bilayer graphene},\ }\href
  {https://doi.org/10.1140/epjst/e2007-00229-1} {\bibfield  {journal} {\bibinfo
   {journal} {The European Physical Journal Special Topics}\ }\textbf {\bibinfo
  {volume} {148}},\ \bibinfo {pages} {91} (\bibinfo {year} {2007})}\BibitemShut
  {NoStop}%
\bibitem [{\citenamefont {Kohn}\ and\ \citenamefont
  {Luttinger}(1955)}]{PhysRev.98.915}%
  \BibitemOpen
  \bibfield  {author} {\bibinfo {author} {\bibfnamefont {W.}~\bibnamefont
  {Kohn}}\ and\ \bibinfo {author} {\bibfnamefont {J.~M.}\ \bibnamefont
  {Luttinger}},\ }\bibfield  {title} {\bibinfo {title} {Theory of donor states
  in silicon},\ }\href {https://doi.org/10.1103/PhysRev.98.915} {\bibfield
  {journal} {\bibinfo  {journal} {Phys. Rev.}\ }\textbf {\bibinfo {volume}
  {98}},\ \bibinfo {pages} {915} (\bibinfo {year} {1955})}\BibitemShut
  {NoStop}%
\bibitem [{\citenamefont {Kohn}(1957)}]{KOHN1957257}%
  \BibitemOpen
  \bibfield  {author} {\bibinfo {author} {\bibfnamefont {W.}~\bibnamefont
  {Kohn}},\ }\href
  {https://doi.org/https://doi.org/10.1016/S0081-1947(08)60104-6} {\emph
  {\bibinfo {title} {Shallow Impurity States in Silicon and Germanium}}},\
  edited by\ \bibinfo {editor} {\bibfnamefont {F.}~\bibnamefont {Seitz}}\ and\
  \bibinfo {editor} {\bibfnamefont {D.}~\bibnamefont {Turnbull}},\ \bibinfo
  {series} {Solid State Physics}, Vol.~\bibinfo {volume} {5}\ (\bibinfo
  {publisher} {Academic Press},\ \bibinfo {year} {1957})\ pp.\ \bibinfo {pages}
  {257--320}\BibitemShut {NoStop}%
\bibitem [{\citenamefont {Bassani}\ \emph {et~al.}(1974)\citenamefont
  {Bassani}, \citenamefont {Iadonisi},\ and\ \citenamefont
  {Preziosi}}]{Bassani1974}%
  \BibitemOpen
  \bibfield  {author} {\bibinfo {author} {\bibfnamefont {F.}~\bibnamefont
  {Bassani}}, \bibinfo {author} {\bibfnamefont {G.}~\bibnamefont {Iadonisi}},\
  and\ \bibinfo {author} {\bibfnamefont {B.}~\bibnamefont {Preziosi}},\
  }\bibfield  {title} {\bibinfo {title} {Electronic impurity levels in
  semiconductors},\ }\href {https://doi.org/10.1088/0034-4885/37/9/001}
  {\bibfield  {journal} {\bibinfo  {journal} {Reports on Progress in Physics}\
  }\textbf {\bibinfo {volume} {37}},\ \bibinfo {pages} {1099} (\bibinfo {year}
  {1974})}\BibitemShut {NoStop}%
\bibitem [{\citenamefont {Shklovskii}\ and\ \citenamefont
  {Efros}(1984)}]{Shklovskii2013}%
  \BibitemOpen
  \bibfield  {author} {\bibinfo {author} {\bibfnamefont {B.~I.}\ \bibnamefont
  {Shklovskii}}\ and\ \bibinfo {author} {\bibfnamefont {A.~L.}\ \bibnamefont
  {Efros}},\ }\href@noop {} {\emph {\bibinfo {title} {Electronic properties of
  doped semiconductors}}},\ \bibinfo {series} {Springer series in solid-state
  sciences}\ No.~\bibinfo {number} {45}\ (\bibinfo  {publisher}
  {Springer-Verlag, Berlin},\ \bibinfo {address} {Berlin},\ \bibinfo {year}
  {1984})\BibitemShut {NoStop}%
\bibitem [{rem()}]{remark}%
  \BibitemOpen
  \href@noop {} {}\bibinfo {note} {In the dynamical case where the polarization
  function depends on frequency, the screened Coulomb interaction between
  quasiparticles can affect the plasmon frequency}\BibitemShut {NoStop}%
\bibitem [{\citenamefont {Prudnikov}\ \emph {et~al.}(2002)\citenamefont
  {Prudnikov}, \citenamefont {Brychkov},\ and\ \citenamefont
  {Marichev}}]{PrudnikovII}%
  \BibitemOpen
  \bibfield  {author} {\bibinfo {author} {\bibfnamefont {A.~P.}\ \bibnamefont
  {Prudnikov}}, \bibinfo {author} {\bibfnamefont {Y.~A.}\ \bibnamefont
  {Brychkov}},\ and\ \bibinfo {author} {\bibfnamefont {O.~I.}\ \bibnamefont
  {Marichev}},\ }\href@noop {} {\emph {\bibinfo {title} {Integrals and
  {S}eries. {S}pecial {F}unctions}}},\ Vol.~\bibinfo {volume} {2}\ (\bibinfo
  {publisher} {Taylor \& Francis, London},\ \bibinfo {year} {2002})\BibitemShut
  {NoStop}%
\bibitem [{\citenamefont {Rytova}(1967)}]{Rytova}%
  \BibitemOpen
  \bibfield  {author} {\bibinfo {author} {\bibfnamefont {N.~S.}\ \bibnamefont
  {Rytova}},\ }\bibfield  {title} {\bibinfo {title} {The screened potential of
  a point charge in a thin film},\ }\href@noop {} {\bibfield  {journal}
  {\bibinfo  {journal} {Moscow Univ. Phys. Bull.}\ }\textbf {\bibinfo {volume}
  {3}},\ \bibinfo {pages} {18} (\bibinfo {year} {1967})}\BibitemShut {NoStop}%
\bibitem [{\citenamefont {Chaplik}\ and\ \citenamefont
  {\'{E}ntin}()}]{Chaplik}%
  \BibitemOpen
  \bibfield  {author} {\bibinfo {author} {\bibfnamefont {A.~V.}\ \bibnamefont
  {Chaplik}}\ and\ \bibinfo {author} {\bibfnamefont {M.~V.}\ \bibnamefont
  {\'{E}ntin}},\ }\bibfield  {title} {\bibinfo {title} {Charged impurities in
  very thin layers},\ }\href@noop {} {\bibinfo  {journal} {Zh. Eksp. Teor. Fiz.
  61, 2496 (1971) [J.Exp. Theor. Phys. 34, 1335 (1971)]}\ }\BibitemShut
  {NoStop}%
\bibitem [{\citenamefont {Keldysh}()}]{Keldysh}%
  \BibitemOpen
\bibfield  {journal} {  }\bibfield  {author} {\bibinfo {author} {\bibfnamefont
  {L.~V.}\ \bibnamefont {Keldysh}},\ }\bibfield  {title} {\bibinfo {title}
  {Coulomb interaction in thin semiconductor and semimetal films},\ }\href@noop
  {} {\bibinfo  {journal} {Pis\'{}ma Zh. Eksp. Teor. Fiz. 29, 716 (1979) [Sov.
  J. Exp.Theor. Phys. Lett. 29, 658 (1979)]}\ }\BibitemShut {NoStop}%
\bibitem [{\citenamefont {Kondovych}\ \emph {et~al.}(2017)\citenamefont
  {Kondovych}, \citenamefont {Luk\'{}yanchuk}, \citenamefont {Baturina},\ and\
  \citenamefont {Vinokur}}]{Vinokur2017}%
  \BibitemOpen
\bibfield  {journal} {  }\bibfield  {author} {\bibinfo {author} {\bibfnamefont
  {S.}~\bibnamefont {Kondovych}}, \bibinfo {author} {\bibfnamefont
  {I.}~\bibnamefont {Luk\'{}yanchuk}}, \bibinfo {author} {\bibfnamefont
  {T.~I.}\ \bibnamefont {Baturina}},\ and\ \bibinfo {author} {\bibfnamefont
  {V.~M.}\ \bibnamefont {Vinokur}},\ }\bibfield  {title} {\bibinfo {title}
  {Gate-tunable electron interaction in high-$\kappa$ dielectric films},\
  }\href {https://doi.org/10.1038/srep42770} {\bibfield  {journal} {\bibinfo
  {journal} {Scientific Reports}\ }\textbf {\bibinfo {volume} {7}},\ \bibinfo
  {pages} {42770} (\bibinfo {year} {2017})}\BibitemShut {NoStop}%
\bibitem [{\citenamefont {Van~Pottelberge}\ \emph {et~al.}(2018)\citenamefont
  {Van~Pottelberge}, \citenamefont {Zarenia},\ and\ \citenamefont
  {Peeters}}]{Peeters}%
  \BibitemOpen
  \bibfield  {author} {\bibinfo {author} {\bibfnamefont {R.}~\bibnamefont
  {Van~Pottelberge}}, \bibinfo {author} {\bibfnamefont {M.}~\bibnamefont
  {Zarenia}},\ and\ \bibinfo {author} {\bibfnamefont {F.~M.}\ \bibnamefont
  {Peeters}},\ }\bibfield  {title} {\bibinfo {title} {Magnetic field dependence
  of atomic collapse in bilayer graphene},\ }\href
  {https://doi.org/10.1103/physrevb.98.115406} {\bibfield  {journal} {\bibinfo
  {journal} {Physical Review B}\ }\textbf {\bibinfo {volume} {98}},\ \bibinfo
  {pages} {115406} (\bibinfo {year} {2018})}\BibitemShut {NoStop}%
\bibitem [{\citenamefont {Fl\"{u}gge}(1999)}]{Flugge}%
  \BibitemOpen
  \bibfield  {author} {\bibinfo {author} {\bibfnamefont {S.}~\bibnamefont
  {Fl\"{u}gge}},\ }\href@noop {} {\emph {\bibinfo {title} {Practical quantum
  mechanics}}},\ Vol.\ \bibinfo {volume} {1, Problems}\ (\bibinfo  {publisher}
  {Springer, Berlin},\ \bibinfo {year} {1999})\BibitemShut {NoStop}%
\end{thebibliography}%

\end{document}